\documentclass[12pt]{article}

\usepackage{amsmath}

\newtheorem{proposition}{Proposition}

\begin{document}

\title{A peeling theorem  for the Weyl tensor in higher dimensions}

\author{Selim Amar}

\maketitle
\noindent
Email: selim.amar@mail.mcgill.ca \\
Department of Mathematics and Statistics, McGill University, 805 Sherbrooke St. West, Montréal H3A 0B9, Québec, Canada

\vspace{10pt}

\begin{abstract}
A peeling theorem for the Weyl tensor in higher dimensional Lorentzian manifolds is presented. We obtain it by generalizing a proof from the four dimensional case. We derive a generic behavior, discuss interesting subcases and retrieve the four dimensional result.
\end{abstract}

\section{Introduction}

The peeling theorem is a result in general relativity describing the asymptotic decay of the Weyl tensor in an asymptotically flat vacuum spacetime (see \cite{Bondi_Peeling}, \cite{Sachs_Peeling} and \cite{NP_Peeling} for the original theorem). In terms of the Petrov classification based on boost weights (see \cite{MILSON_2005}, \cite{Coley_2004}, \cite{Coley_2008} and \cite{Ortaggio_2012} for this approach to classify the Weyl tensor), the theorem states that the Weyl scalars of boost weight $w$ decay like $O(r^{-3-w})$. The approach taken in \cite{NP_Peeling} to prove it is to use the Bianchi and Ricci identities (along with commutators of covariant derivatives) to derive ordinary differential equations for the scalars $\Psi_i$'s. A simple proposition is then used recursively to find the decay of these quantities with the initial assumption $\Psi_0 = O(r^{-5})$ being used to start things off. This method is generalized here to higher dimensions. We will work with a null tetrad $ \mathbf{m}_0 = \boldsymbol{\ell}, \mathbf{m}_1 = \mathbf{n}, \mathbf{m}_i$ whose only nonzero inner products are $g(\boldsymbol{\ell}, \mathbf{n}) = 1$ and $g(\mathbf{m}_i, \mathbf{m}_j) = \delta_{ij}$. The indices $i, j, k, l$ range from $2$ to $d-1$ while $a, b, c, e$ range from $0$ to $d-1$. Instead of finding the rates of decay of the components of the Weyl tensor directly, we'll instead use their decompositions under the action of the $(d-2)$ orthogonal group acting on the $\mathbf{m}_i$'s. The derivation of these decompositions and their use in classifying the Weyl tensor in higher dimension were described in  \cite{Coley_2009} and \cite{Coley_2012}. This is done in order to have a more refined peeling theorem since the resulting irreducible components can potentially decay with different rates. If we denote by $C_{abce} = C(\mathbf{m}_{a}, \mathbf{m}_{b}, \mathbf{m}_{c}, \mathbf{m}_{e})$ the components of the Weyl tensor in the null tetrad then their irreducible components are defined by the following expansions (they also have some specific algebraic properties; see \cite{Coley_2009} for further details):

\begin{table}[hbt!]
\caption{Irreducible components of the Weyl scalars.}
    \label{tab:irr_comp}
    \centering
    \renewcommand{\arraystretch}{1.5}
    \begin{tabular}{|c|c|c|}
    \hline
         Boost Weight & Irreducible Components & Weyl Scalars\\
         \hline
         +2 & $\hat{H}_{ij}$ & $C_{0i0j} = \hat{H}_{ij}$ \\
         +1 & $\hat{v}_{i}, \hat{T}_{ijk}$ & $C_{0ijk} = \delta_{ij}\hat{v}_k - \delta_{ik}\hat{v}_{j} + \hat{T}_{ijk}$ \\ & & $C_{010i} = -(d-3)\hat{v}_{i}$ \\
         0 & $A_{ij}, \Bar{R}, \Bar{S}_{ij}, \Bar{C}_{ijkl}$ & $C_{01ij} = A_{ij}$   $C_{0101} = -\frac{1}{2}\Bar{R}$ \\
         & & $C_{0i1j} = \frac{1}{2}A_{ij} - \frac{1}{2}\Bar{R}_{ij}$\\
         & & $C_{ijkl} = \Bar{C}_{ijkl} + \frac{2}{d-4}(\delta_{i[k}\Bar{R}_{l]j} - \delta_{j[k}\Bar{R}_{l]i})$ \\ 
         & & $- \frac{2}{(d-3)(d-4)} \Bar{R} \delta_{i[k} \delta_{l]j}$ \\
         & & $C^{k}\!_{ikj} = \Bar{R}_{ij} = \Bar{S}_{ij} + \frac{1}{d-2}\Bar{R}\delta_{ij}$ \\
         -1 & $\check{v}_{i}, \check{T}_{ijk}$ & $C_{1ijk} = \delta_{ij}\check{v}_k - \delta_{ik}\check{v}_{j} + \check{T}_{ijk}$ \\
         & & $C_{101i} = -(d-3)\check{v}_{i}$ \\
         -2 & $\check{H}_{ij}$ & $C_{1i1j} = \check{H}_{ij}$\\
    \hline
    \end{tabular}
\end{table}

The coordinate $r$ will denote an affine parameter along the geodesics of $\boldsymbol{\ell}$ and $x^{A}$ will denote the other coordinates. Then, assuming $\hat{H}_{ij} = O(r^{-\nu}); \nu \geq 4$ along with other technicalities, the generic decay of the irreducible components found is:

\begin{equation}
\begin{aligned}
    \hat{v}_{i}, \hat{T}_{ijk} = O(r^{-3}) \quad \Bar{R}, \Bar{S}_{ij}, \Bar{C}_{ijkl}, A_{ij} = O(r^{-2}) \\
    \check{v}_{i}, \check{T}_{ijk} = O(r^{-1}) \quad \check{H}_{ij} = O(1)
\end{aligned}
\end{equation}

 This generic decay is slower than what was derived in \cite{Godazgar_2012} and \cite{Ortaggio_2014} but complements these results by having a weaker hypothesis (in the Ricci-flat case), by using a different method and by working with a refined Petrov classification based on the irreducible components of the Weyl scalars. The generic behavior is derived in section \ref{subsec1} while various subcases are discussed in section \ref{subsec2} along with how to retrieve the four dimensional result.

\section{Derivation}

The result we'll use is the following (see \cite{NP_Peeling} for the proof):

\begin{proposition}
Let A be a constant n$\times$n complex matrix whose eigenvalues have nonpositive real part. Let $B(r)$ be a complex n$\times$n matrix and $b(r)$ be a vector satisfying:
\begin{equation}
    B = O(r^{-2}) \quad b = O(r^{-2})
\end{equation}
Then, the solutions to
\begin{equation}
    \frac{dy}{dr} = (\frac{A}{r} + B)y + b
\end{equation}
are all bounded as $r \longrightarrow \infty$.
\label{proposition}
\end{proposition}

Note that if $\frac{\partial B}{\partial x^{A}}$ and $\frac{\partial b}{\partial x^{A}}$ are still $O(r^{-2})$, we can differentiate with respect to $x^{A}$ on each side and apply the proposition once more to conclude that $\frac{\partial y}{\partial x^{A}}$ is bounded. This fact will be used again and again to get the behavior of the $D^{\alpha}$ derivatives of $y$ \footnote{$D^{\alpha}$ will denote an arbitrary derivative of order smaller than or equal to $\alpha$ with respect to the $x^{A}$ coordinates (mixed derivatives are allowed).}. Other quantities we'll find the decay of are the Ricci rotation coefficients $L_{ab}, N_{ab}$ and $\overset{i}{M}_{ab}$:

\begin{equation}
    \nabla \boldsymbol{\ell} ^{\flat}  = L_{ab}\mathbf{m}^{a}\otimes \mathbf{m}^b \quad \nabla \mathbf{n}^{\flat} =  N_{ab}\mathbf{m}^{a}\otimes \mathbf{m}^{b} \quad \nabla \mathbf{m}_i^{\flat} = \overset{i}{M}_{ab}\mathbf{m}^{a}\otimes \mathbf{m}^{b}
\end{equation}
And the coefficients of the null tetrad in the coordinate basis:
\begin{equation}
    \boldsymbol{\ell} = \frac{\partial}{\partial r} \quad \mathbf{n} = U \frac{\partial}{\partial r} + X^{A}\frac{\partial}{\partial x^{A}} \quad \mathbf{m}_{i} = \omega_{i}\frac{\partial}{\partial r} + \xi^{A}_{i}\frac{\partial}{\partial x^{A}}
\end{equation}. 

The Bianchi identities will give us the $r$-derivatives of the irreducible components of the Weyl scalars while the Ricci identities and commutator relations will give us the ones of the Ricci rotation coefficients and of the coefficients of the tetrad. These identities can be found in \cite{Bianchi_Identities}, \cite{Ricci_Identities} and \cite{Commutators}. We'll be assuming the decay rate of the Weyl scalar of boost weight +2 and it's derivatives up to order 4, that the tetrad frame is parallel transported along $\boldsymbol{\ell}$ and that the optical matrix $L_{ij}$ is asymptotically non-singular and has a non-zero $\frac{1}{r}$ term. Furthermore, we assume that the Ricci tensor vanishes. This amounts to

\begin{equation}
    \hat{H}_{ij} = O(r^{-\nu}) \quad D^{4}\hat{H}_{ij} = O(r^{-\nu}) \quad D^{3} \frac{\partial \hat{H}_{ij}}{\partial r} = O(r^{-\nu - 1}); \nu \geq 4
\end{equation}
\begin{equation}
    L_{a0}, N_{a0}, \overset{a}{M}_{b0} = 0 \quad R_{ab} = 0 \quad
    L_{ij} = \frac{\delta_{ij}}{r} + O(r^{-2}) 
\end{equation}
where $L_{ij}$ can be found from the Sachs equation
\begin{equation}
    \frac{\partial L_{ij}}{\partial r} = -L_{ik}L_{kj} - \hat{H}_{ij}
\end{equation}

To do so, we can proceed exactly as is done in \cite{NP_Peeling} for the four dimensional case since all steps are valid in higher dimensions. We start by solving the system

\begin{equation}
    \frac{\partial Y_{ij}}{\partial r} = L_{ik}Y_{kj}
\end{equation}
It has the following solution ($F$ and $E$ are two constants matrices fixed by initial conditions; see \cite{NP_Peeling}):
\begin{equation}
    \frac{\partial Y_{ij}}{\partial r} = F_{ij} + O(r^{-\nu+2})
\end{equation}
\begin{equation}
    Y_{ij} = rF_{ij} + E_{ij} + O(r^{-\nu+3})
\end{equation}

Then, $L = \frac{\partial Y}{\partial r}Y^{-1}$ is of the form $L_{ij} = \frac{\delta_{ij}}{r} + O(r^{-2})$ if F is invertible while it is either asymptotically singular or with no $\frac{1}{r}$ term if F is not invertible.

\subsection{Generic Parallel Transported Frame}\label{subsec1}

We must first find the behavior of the derivatives of $L_{ij}$. By differentiating the Sachs equation once and using $L_{ij} = \frac{\delta_{ij}}{r} + O(r^{-2})$ we find:

\begin{equation}
    \frac{\partial (r^2 D^{1}L_{ij})}{\partial r} = - O_{ik}(r^{-2})(r^2 D^{1}L_{kj}) - (r^2 D^{1}L_{ik})O_{kj}(r^{-2}) -  r^2 D^{1}\hat{H}_{ij}
\end{equation}

We can apply proposition \ref{proposition} to this system. We thus have $D^{1}L_{ij} = O(r^{-2})$. Similarly, we can differentiate 3 more times with respect to $x^{A}$ and apply proposition \ref{proposition} repeatedly to find $D^{4}L_{ij} = O(r^{-2})$. To obtain the asymptotics of the irreducible components, we start with the decay rate of $\hat{v}$ and $\hat{T}$. By computing the irreducible parts of (B.8) \footnote{(B.\#) refer to the Bianchi equations in \cite{Bianchi_Identities}.} we can apply proposition \ref{proposition} to the following system:

\begin{equation}
\frac{\partial}{\partial r}
    \begin{pmatrix} 
    \omega_{i}\\
    r \xi^{A}_i \\
    rL_{1i} \\
    r\overset{i}{M}_{jk} \\
    r^3\hat{v}_i \\
    r^3\hat{T}_{ijk}
    \end{pmatrix}
    = \left( \begin{pmatrix}
    -1 & 0 & -1 & 0 & 0 & 0 \\
    0 & 0 & 0 & 0 & 0 & 0 \\
    0 & 0 & 0 & 0 & 0 & 0 \\
    0 & 0 & 0 & 0 & 0 & 0 \\
    0 & 0 & 0 & 0 & -(d-3)& 0\\
    0 & 0 & 0 & 0 & 0 & 0\\
    \end{pmatrix} \frac{1}{r}
    + B_1 \right)
    \begin{pmatrix}
    \omega_{i}\\
    r \xi^{A}_i \\
    rL_{1i} \\
    r\overset{i}{M}_{jk} \\
    r^3\hat{v}_i \\
    r^3\hat{T}_{ijk}
    \end{pmatrix}
    + b_1
\end{equation}

Where $B_1, b_1$ satisfy $B_1, b_1 = O(r^{-2})$ and are composed of certain expressions in $L_{ij} - \frac{\delta_{ij}}{r}$ together with $\hat{H}_{ij}$ and it's derivatives. We cannot obtain a faster rate of decay for $\hat{v}_{i}$ since the off-diagonal terms in $B_1$ coupling $\hat{T}_{ijk}$ and $\hat{v}_{i}$ would not be $O(r^{-2})$ if we used $r^{\beta}\hat{v}_{i}$ ($\beta>3$) instead of $r^{3}\hat{v}_{i}$. Such coupling preventing us from deriving different rates of decay is present for all boost weights and is the reason we cannot prove peeling between different irreducible components of equal boost weight in the generic case. We thus find

\begin{equation}
    \omega_i = O(1) \quad \overset{i}{M}_{jk}, L_{1i}, \xi^{A}_i = O(r^{-1}) \quad \hat{v}_i, \hat{T}_{ijk} = O(r^{-3})
\end{equation}
We can plug this back in equation (B.8) and get
\begin{equation}
    \frac{\partial \hat{v}_i}{\partial r}, \frac{\hat{T}_{ijk}}{\partial r} = O(r^{-4})
\end{equation}

Furthermore, we can differentiate this system three times with respect to $x^{A}$ and still apply proposition \ref{proposition}. Thus, all the $D^{3}$ derivatives of these quantities have the same decay. We now move on to components of boost weight 0. We use (B.3), (B.5) and (B.12) to derive:

\begin{equation}
\begin{aligned}
\frac{\partial}{\partial r}
    \begin{pmatrix} 
    r^{-1}U\\
    L_{11} \\
    \overset{i}{M}_{j1}\\
    X^{A} \\
    rN_{ij} \\
    rL_{i1} \\
    r^2 A_{ij} \\
    r^2 \Bar{R}\\
    r^2 \Bar{S}_{ij}\\
    r^2 \Bar{C}_{ijkl}
    \end{pmatrix}
    = \left( \frac{A_0}{r}
    + B_0 \right)
    \begin{pmatrix} 
    r^{-1}U\\
    L_{11} \\
    \overset{i}{M}_{j1}\\
    X^{A} \\
    rN_{ij} \\
    rL_{i1} \\
    r^2 A_{ij} \\
    r^2 \Bar{R}\\
    r^2 \Bar{S}_{ij}\\
    r^2 \Bar{C}_{ijkl}
    \end{pmatrix}
    + b_0 
    \hspace{2.5em} \\ 
    A_0 = \begin{pmatrix}
    0 & -1 & 0 & 0 & 0 & 0 & 0 & 0 & 0 & 0 \\
    0 & 0 & 0 & 0 & 0 & 0 & 0 & 0 & 0 & 0 \\
    0 & 0 & 0 & 0 & 0 & 0 & 0 & 0 & 0 & 0 \\
    0 & 0 & 0 & 0 & 0 & 0 & 0 & 0 & 0 & 0 \\
    0 & 0 & 0 & 0 & 0 & 0 & 0 & 0 & 0 & 0 \\
    0 & 0 & 0 & 0 & 0 & 0 & 0 & 0 & 0 & 0 \\
    0 & 0 & 0 & 0 & 0 & 0 & -1 & 0 & 0 & 0 \\
    0 & 0 & 0 & 0 & 0 & 0 & 0 & -(d-3) & 0 & 0 \\
    0 & 0 & 0 & 0 & 0 & 0 & 0 & 0 & -(d/2-2) & 0 \\
    0 & 0 & 0 & 0 & 0 & 0 & 0 & 0 & 0 & 0 \\
    \end{pmatrix}
\end{aligned}
\end{equation}

$B_0, b_0$ are given in terms of the quantities whose asymptotics we know from above (and are $O(r^{-2})$). This time, the rate of decay is capped by $b_0$ due to $\hat{T}_{ijk}, \hat{v}_i$ decaying only like $O(r^{-3})$. Thus, we have

\begin{equation}
\begin{aligned}
    U = O(r) \quad L_{11}, \overset{i}{M}_{j1}, X^{A} = O(1) \quad N_{ij}, L_{i1} = O(r^{-1}) \\ A_{ij}, \Bar{R}, \Bar{S}_{ij}, \Bar{C}_{ijkl} = O(r^{-2})
\end{aligned}
\end{equation}

Again, we can put back these results into (B.3), (B.5) and (B.12) to get the decay of the $\frac{\partial}{\partial r}$ derivatives:

\begin{equation}
    \frac{\partial A_{ij}}{\partial r}, \frac{\partial \Bar{R}}{\partial r}, \frac{\partial \Bar{S}_{ij}}{\partial r}, \frac{\partial \Bar{C}_{ijkl}}{\partial r} = O(r^{-3})
\end{equation}

We can differentiate the system twice with respect to $x^{A}$ and apply proposition \ref{proposition} to get that the $D^{2}$ derivatives of all these quantities behave the same way. Next are the components of boost weight -1. Their derivatives are found from (B.6) and (B.9). The system is as follows:

\begin{equation}
\frac{\partial}{\partial r}
    \begin{pmatrix} 
    r^{-1}N_{i1} \\
    r \check{v}_{i} \\
    r \check{T}_{ijk}
    \end{pmatrix}
    = \left( \begin{pmatrix}
    0 & 0 & 0 \\
    0 & -(d-3) & 0 \\
    0 & 0 & -1 \\
    \end{pmatrix} \frac{1}{r}
    + B_{-1} \right)
    \begin{pmatrix}
    r^{-1}N_{i1} \\
    r \check{v}_{i} \\
    r \check{T}_{ijk}
    \end{pmatrix}
    + b_{-1}
\end{equation}

The decay rate is capped by the one of the components of boost weight 0. Applying proposition \ref{proposition} gives us the decay of $N_{i1}, \hat{v}_i, \hat{T}_{ijk}$ and just like before we can plug the results back into (B.6) and (B.9) to get the behavior of their $\frac{\partial}{\partial r}$ derivative and we can differentiate the system once with respect to $x^{A}$ to get the behavior of their $D^1$ derivatives. We thus have:

\begin{equation}
\begin{aligned}
    N_{i1} = O(r) \quad \check{v}_i, \check{T}_{ijk} = O(r^{-1}) \\
    \frac{\partial \check{v}_i}{\partial r}, \frac{\partial \check{T}_{ijk}}{\partial r} = O(r^{-2})
\end{aligned}
\end{equation}
With their $D^1$ derivatives behaving the same way. Finally, we find the decay rate of $\check{H}_{ij}$. We use equation (B.4) with $A=0$:

\begin{equation}
\frac{\partial \check{H}_{ij}}{\partial r}
    = -\frac{\check{H}_{ij}}{r} + O(r^{-2})\check{H}_{ij} + O(r^{-2})
\end{equation}

This gives us

\begin{equation}
    \check{H}_{ij} = O(1)
\end{equation}

\subsection{Subcases and Four Dimensional Result}\label{subsec2}

The irreducible components of equal boost weight decouple from each other when the $B_{w}$'s no longer mixes them. This happens when certain contractions between these components and $L_{ij} - \frac{\delta_{ij}}{r}$ vanish. In that case, the decay is only limited by the condition that $A$ has nonpositive eigenvalues and the decays of the scalars of bigger boost weights. The exponents are then:

\begin{table}[hbt!]
\renewcommand{\arraystretch}{1.5}
    \centering
    \begin{tabular}{c|c|c|c|c}
    $\hat{H}_{ij}$ & $\hat{v}_{i}$ & $\hat{T}_{ijk}$ & $\Bar{R}$ & $\Bar{S}_{ij}$ \\
    $\nu$ & $min(d, \nu-1)$ & $3$ & $min(d-1, \nu-2)$ & $2$ \\
    \hline
    $\Bar{C}_{ijkl}$ & $A_{ij}$ & $\check{v}_{i}$ & $\check{T}_{ijk}$ & $\check{H}_{ij}$ \\
    $2$ & $min(3, \nu-2)$ & $1$ & $1$ & $0$ \\
       
    \end{tabular}
\end{table}

\pagebreak

Here, the rates of decay of $\Bar{S}_{ij}$, $\check{v}_i$, $\check{T}_{ijk}$ and $\check{H}_{ij}$ are still limited by the one of $\hat{T}_{ijk}$ since the latter decays too slowly at a rate of $O(r^{-3})$. If we furthermore assume that $\Bar{S}_{ij}$, $\check{v}_i$ and $\check{H}_{ij}$ decouple from $\hat{T}_{ijk}$ we find:

\begin{table}[hbt!]
\renewcommand{\arraystretch}{1.5}
    \centering
    \begin{tabular}{c|c|c|c|c}
    $\hat{H}_{ij}$ & $\hat{v}_{i}$ & $\hat{T}_{ijk}$ & $\Bar{R}$ & $\Bar{S}_{ij}$ \\
    $\nu$ & $min(d, \nu-1)$ & $3$ & $min(d-1, \nu-2)$ & $min(d/2, v-2)$ \\
    \hline
    $\Bar{C}_{ijkl}$ & $A_{ij}$ & $\check{v}_{i}$ & $\check{T}_{ijk}$ & $\check{H}_{ij}$ \\
    $2$ & $min(3, \nu-2)$ & $min(d-2, \nu-3)$ & $min(2, \nu-3)$ & $min(1, v-4)$ \\
       
    \end{tabular}
\label{tab}
\end{table}

To retrieve the four dimensional case, note that in four dimensions $\hat{T}_{ijk}$, $\Bar{S}_{ij}$, $\Bar{C}_{ijkl}$ and $\check{T}_{ijk}$ vanish. Thus, we find ourselves in the subcase just described hence the decay is given by the table above with $d=4$ and $\nu = 5$ (while $A_{ij}$ and $\Bar{R}$ might not decouple, they have the same rate of decay for $d=4$). Various other subcases can be derived by assuming that certain irreducible components decouple from each other in the Bianchi identities \cite{Bianchi_Identities} and following the steps laid out in section \ref{subsec1} with these additional assumptions. This allows for more opportunity of peeling. One such subcase of interest is when all the irreducible components decouple from each other and the rates of decay we can predict are the maximum allowed by the restriction on $A$:

\begin{table}[hbt!]
\renewcommand{\arraystretch}{1.5}
    \centering
    \begin{tabular}{c|c|c|c|c}
    $\hat{H}_{ij}$ & $\hat{v}_{i}$ & $\hat{T}_{ijk}$ & $\Bar{R}$ & $\Bar{S}_{ij}$ \\
    $\nu$ & $d$ & $3$ & $d-1$ & $d/2$ \\
    \hline
    $\Bar{C}_{ijkl}$ & $A_{ij}$ & $\check{v}_{i}$ & $\check{T}_{ijk}$ & $\check{H}_{ij}$ \\
    $2$ & $3$ & $d-2$ & $2$ & $1$ \\
     
    \end{tabular}
\end{table}

\section{Conclusion}\label{sec1}

In conclusion, we've generalize the use of proposition \ref{proposition} to derive various possible decays for the irreducible components of the Weyl scalars in arbitrary dimension. We've found that in the generic case, the decay that we can predict is independent of the dimension $d$ and of the rate of decay of $\hat{H}_{ij}$ (as long as it's at least $O(r^{-4})$). No peeling between the different irreducible components of equal boost weight can be shown to take place (with this method) in the generic case. This suggest that in the class of Lorentzian manifolds determined by our assumptions, the dimension of the manifold and it's asymptotic properties (as determined by $\nu$) might have no impact on the decay of it's Weyl tensor (generically). Furthermore, for such a generic Weyl tensor, the rates of decay of it's irreducible components only depend on their boost weight. They thus 'peel off' only in groups of components with the same boost weight and the behavior is similar to the one in four dimensions, albeit with slightly slower rates of decay. It would be interesting to determine whether Lorentzian manifolds satisfying our assumptions and whose Weyl tensor has the generic decay derived here exist for all dimensions. This could answer whether the class of manifolds considered here is different from the ones determined by the assumptions of \cite{Godazgar_2012} and \cite{Ortaggio_2014} and could confirm that the assumptions made in \cite{Godazgar_2012} and \cite{Ortaggio_2014} are necessary to obtain the stronger decays they derived. Furthermore, we've mentioned several subcases with interesting peeling between the different irreducible components, one of which giving us the usual four dimensional result. These subcases are reminiscent of the ones derived in \cite{Ortaggio_2014} but are defined by different assumptions. This peeling between irreducible components is an interesting phenomenon from higher dimensions which is due to the richer algebraic structure of the Weyl tensor there. It tells us that in specific cases the decay of the Weyl tensor can indeed depend on the properties of the Lorentzian manifold (as was found in \cite{Godazgar_2012} and \cite{Ortaggio_2014}) but also on the coupling between different irreducible components. For such special cases, the irreducible components 'peel off' in different groups than the ones determined by their boost weights and the behavior differs from the four dimensional case. Just as in four dimensions, we can also ask what would be the corresponding results in the Einstein-Maxwell theory. Furthermore, generalizing the steps to allow for a nonzero Ricci tensor (for example to include a cosmological constant) would also be an interesting route to follow.

\section{Acknowledgments}

I first want to thank Professor Niky Kamran for suggesting this topic and supervising this research. His advices and guidance were critical to the project and have made the learning experience very pleasant and fruitful. Furthermore, I am grateful to the Natural Sciences and Engineering Research Council of Canada for financing this summer's work.

We acknowledge the support of the Natural Sciences and Engineering Research Council of Canada (NSERC).

Nous remercions le Conseil de recherches en sciences naturelles et en génie du Canada (CRSNG) de son soutien.

\bibliographystyle{acm}
\bibliography{main}

\end{document}